\documentclass[twocolumn,secnumarabic,amssymb,nobibnotes,aps,pra]{revtex4-1}
\usepackage{graphicx,bm,amsmath}
\begin{document}


\title{Factoring 51 and 85 with 8 qubits}

\author{Michael R. Geller and Zhongyuan Zhou}
\affiliation{Department of Physics and Astronomy, University of Georgia, Athens, Georgia 
30602, USA}

\date{\today}

\begin{abstract}
We construct simplified quantum circuits for Shor's order-finding algorithm 
for composites $N$ given by products of the Fermat primes 3, 5, 17, 257, 
and 65537. Such composites, including the previously studied case of 15, 
as well as $51, 85, 771, 1285, 4369, \dots$ have the simplifying property 
that the order of $a$ modulo $N$ for every base $a$ coprime to $N$ is a 
power of 2, significantly reducing the usual phase estimation precision 
requirement. Prime factorization of 51 and 85 can be demonstrated with 
only 8 qubits and a modular exponentiation circuit consisting of no more 
than four {\sf CNOT} gates.
\end{abstract}

\pacs{03.67.Lx} 

\maketitle

\section{Order finding and Fermat primes}

Shor's prime factoring algorithm \cite{ShorSIAMJC97} reduces the factorization of a product $N=pp^\prime$ of distinct odd primes $p$ and $p^\prime$ to that of finding the order $r$ of $a \, {\rm mod} \, N$ for a randomly chosen base $a $ coprime to $N$ (with $1 < a < N$), which can be performed efficiently with a quantum computer. The standard implementation \cite{VandersypenNat01} factors a $b$-bit number with $3b$ qubits using a circuit of depth $O(b^3)$; alternative modular exponentiation circuits can be
used to reduce either the space (qubit number) \cite{BeauregardQIC03} or time \cite{ZalkaPre08} requirements. The case $N\!=\!15,$ which has the simplifying property that all orders are powers of 2, has been demonstrated experimentally by several groups \cite{VandersypenNat01,LuPRL07,LanyonPRL07,PolitiSci09,LuceroNatPhys12}. Recent experiments have also factored $N\!=\!21$ \cite{PengPRL08,MartinLopezNatPhoton12} and $128$ \cite{XuPRL12}.

In this paper we consider the application of Shor's algorithm to products of special primes of the form
\begin{equation}
p_k \equiv 2^{2^k} + 1 \ \  {\rm with} \ \ k=0,1,2,3,4.
\label{fermat primes formula}
\end{equation}
Explicitly, 
\begin{equation}
p= 3, \, 5, \, 17, \, 257, \  {\rm and} \ 65537.
\label{fermat primes explicit}
\end{equation}
Fermat proposed that numbers of the form $2^{2^k} + 1$ for any $k=0,1,2,\dots,$ (called Fermat numbers) are prime; however it is now known that the Fermat numbers with $5 \le k \le 32$ are not prime, and it is not known whether there are additional primes of this form for larger values of $k$.

Products of the form
\begin{widetext}
\begin{eqnarray}
N &=& p_k p_{k'} = (2^{2^k} + 1)(2^{2^{k'}} + 1) ,
\ \  {\rm with} \ \ k,k' \in \lbrace 0,1,2,3,4 \rbrace
\ \  {\rm and} \ \ k \neq k'  \nonumber \\
&=& 15, 51, 85, 771, 1285, 4369, 196611, 327685, 1114129, \  {\rm and} \  16843009
\label{fermat products}
\end{eqnarray}
have the special property that the order of $a \, {\rm mod} \, N$ for every base $a$ coprime to $N$ is a power of 2. This follows from Euler's theorem,
\begin{equation}
a^{\phi(y)} \, {\rm mod} \, y =1,
\label{euler's theorem}
\end{equation}
where $y$ is a positive integer, $\phi(y)$ is the number of positive integers less than $y$ that are coprime to $y$, and ${\rm gcd}(a,y)=1.$ When $p$ and $p'$ are odd primes, all $pp'-1$ positive integers less than $pp'$ are coprime to $pp'$ except for the $p-1$ multiples of $p'$ and the $p'-1$ multiples of $p$, and these exceptions are distinct, so
\begin{equation}
\phi(pp') = pp'-1 - (p-1) - (p'-1) = (p-1)(p'-1). 
\label{totient function}
\end{equation}
This result also follows from Euler's product formula. Thus,
\begin{equation}
a^{(p-1)(p'-1)} \, {\rm mod} \, pp' =1.
\label{tentative order condition}
\end{equation}
Recall that the order $r$ of $a \, {\rm mod} \, N$ is the smallest positive integer $x$
satisfying $a^x \, {\rm mod} \, N =1$; therefore for a composite of the form 
(\ref{fermat products}),
\begin{equation}
\phi(N) = (p_k-1)(p_{k'}-1) = 2^{2^k + 2^{k'}}
\label{order condition}
\end{equation}
must be a {\it multiple} of $r$. Because $r$ must be an integer, we conclude that for any $1<a<N$ with ${\rm gcd}(a,N)=1$,  $r$ is a power of 2 as well.
\end{widetext}

\section{Space requirements and circuit construction}
\label{space requirements and circuit construction section}

\begin{figure}
\includegraphics[width=8.0cm]{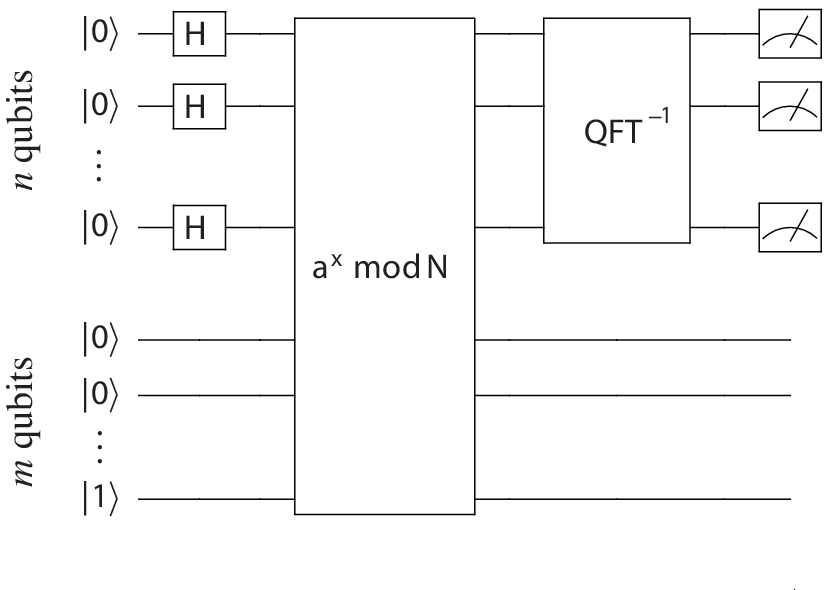} 
\caption{Basic quantum circuit for order finding. Here $n=2b$ and $m=b$, where $b \equiv \lceil \log_2 N \rceil$ is the number of bits in $N$.}
\label{basic circuit figure}
\end{figure} 

The standard \cite{VandersypenNat01} order-finding circuit is shown in Fig.~\ref{basic circuit figure}. The first register has $n$ qubits and the second has $m$. The modular exponentiation operator in Fig.~\ref{basic circuit figure} acts on computational basis states as
\begin{equation}
| x_1 x_2 \cdots x_n \rangle \otimes |0 \cdots 1\rangle \rightarrow | x_1 x_2 \cdots x_n \rangle \otimes | a^x \, {\rm mod} \, N \rangle, 
\label{modular exponentiation operator}
\end{equation}
where
\begin{equation}
x= \sum_{j=1}^n 2^{n-j} \, x_j .
\end{equation}
After the inverse quantum Fourier transform, measurement of the first register is done in the diagonal basis. The probability to observe the value
\begin{equation}
x \in \  \lbrace 0, 1, \dots , 2^n-1 \rbrace
\end{equation}
is
\begin{equation}
{\rm prob} (x) = \frac{\sin^2(\pi r x A/2^n)}{ 2^n \! A \sin^2(\pi r x /2^n)} , 
\label{measurement probability}
\end{equation}
where $r$ is the order and $A$ is the number of distinct values of $x$ such that $a^x \, {\rm mod} \, N$ has the same value (this is approximately $2^n/r$). This probability distribution has peaks at integer values of $x$ near
\begin{equation}
j \times \frac{2^n}{r} \ \  {\rm with} \ \ j=0,1,\cdots,r-1.
\label{peak loctions}
\end{equation}

The number of qubits $n$  in the first register is chosen to enable reliable extraction of the value of $r$ in (\ref{peak loctions}), which depends on whether or not $r$ is a power of 2. In actual applications of Shor's algorithm this will not be known, of course, as the point of the quantum algorithm is to determine $r$. In this usual situation, measurement will yield (with ${\rm prob} > 4/\pi^2$) an $x$ satisfying
\begin{equation}
\bigg| \frac{x}{2^n} - \frac{j}{r} \bigg| \le \frac{1}{2^n} \ \  {\rm with} \ \ j\in\lbrace0,1,\cdots,r-1\rbrace.
\label{n bit measurement condition}
\end{equation}
By choosing $n=2b$ qubits in the first register, where $ b \equiv \lceil \log_2 N \rceil$, we are guaranteed that $j/r$ will be a (continued fraction) convergent  of  $x/2^n.$ However, for the family of composites $N = (2^{2^k} + 1)(2^{2^{k'}} + 1)$ considered here, all bases have orders
\begin{equation}
r = 2^\ell  \ \  {\rm with} \ \ \ell \in \lbrace1,2,3,\cdots, \ell_{\rm max} \rbrace,
\label{order formula}
\end{equation}
where the value of $\ell_{\rm max}$ is discussed below. In this case
\begin{equation}
A = \frac{2^n}{r}
\label{A formula}
\end{equation}
and the peaks (\ref{peak loctions}) in (\ref{measurement probability}) occur
at integral values 
\begin{equation}
x = 0, \ 2^{n-\ell}, \ 2\times 2^{n-\ell}, \cdots , \ (r-1)\times 2^{n-\ell}.
\label{peak values special case}
\end{equation}
Therefore, as long as we have
\begin{equation}
n = \ell_{\rm max}
\label{first register space requirement}
\end{equation}
qubits in the first register we will be able to determine $r$, possibly after a small number of repetitions. The simplest way to extract $r$ from $x$ here (assuming $x\neq 0$) is to simplify the ratio
\begin{equation} 
\frac{x}{2^n}
\end{equation}
down to an irreducible fraction, which will yield both $j$ and $r$ [recall (\ref{peak loctions})] unless they have happen to have a common factor.

Next we discuss the value of $\ell_{\rm max}$ (which determines the largest
order $2^{\ell_{\rm max}}$) for a given composite $N$. We do not have an explicit 
formula for $\ell_{\rm max}$. However, when $N$ is a product of distinct odd primes, 
$r$ can be as large as $\phi(N)/2$ \cite{GerjuoyAJP05}, so for an $N$ of the form (\ref{fermat products})
we have the bound [see (\ref{order condition})]
\begin{equation}
\ell_{\rm max} \le 2^{k} + 2^{k'} -1.
\label{max l condition}
\end{equation}
For example, in the case of $N\!=\!51 \ (k=0, \, k'=2)$, the largest order is $2^4=16$, and the upper bound is realized. However for $N\!=\!85 \ (k=1, \, k'=2)$, it is not (the largest order present is 16, not 32).

The second register stores the values of 
\begin{equation}
a^x \, {\rm mod} \, N \in \lbrace 0,1,\cdots,N-1 \rbrace
\end{equation} 
and therefore normally requires $b$ qubits. However, for a given $a$, only $r$ of these values are distinct. Thus we can use fewer than $b$ qubits. This simplification, while not essential, has been used in all gate-based factoring demonstrations to date. The reduction amounts to computing a table of values of $a^x \, {\rm mod} \, N$ classically for a given base $a$, constructing a corresponding quantum circuit, and ignoring or eliminating unused qubits in the second register. We note that in addition to being unscalable, this method of constructing the modular exponentiation operator implicitly or explicitly uses the value of the order $r$, i.e., the answer which the quantum computation is supposed to determine \cite{usingOrderNote}. We will discuss this issue further in Sec.~\ref{conclusion section}.

In this work we will adopt an equivalent---but perhaps more systematic and transparent---modular exponentiation circuit construction: We follow the output of $a^x \, {\rm mod} \, N$ by a second transformation 
\begin{equation}
\begin{matrix}
1 &&&& 0 \\
a \, {\rm mod} \, N &&&& 1 \\
a^2 \, {\rm mod} \, N && \longrightarrow && 2 \\
\vdots &&&& \vdots \\
a^{r-1} \, {\rm mod} \, N &&&& r-1 \\
\end{matrix},
\label{compression map}
\end{equation}
which maps the $r$ distinct values of $a^x \, {\rm mod} \, N$ to $0, 1, \dots , r-1$. In (\ref{compression map}) we assume that $1 < a < N$. We refer to this classical pre-processing of $a^x \, {\rm mod} \, N$ as {\it compression}. Compression does not adversely affect the operation of the order-finding circuit, but reduces $m$ from $b$ to $\ell_{\rm max}$ in a systematic manner (and generalizes the ``full compilation" method of Ref.~[\onlinecite{LanyonPRL07}].)

\begin{figure}
\includegraphics[width=5.0cm]{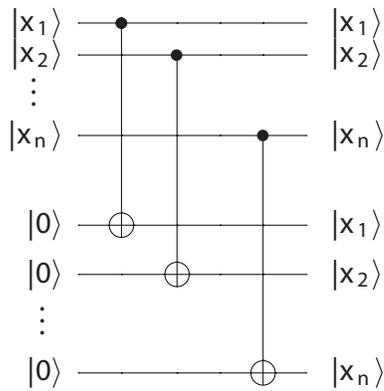} 
\caption{Circuit to copy the first register to the second.}
\label{copy circuit figure}
\end{figure}

Note that any set of $r$ distinct non-negative integers---in any order---could be used for the output of the compression map (\ref{compression map}). However the choice employed here, and indicated in (\ref{compression map}), is especially simple because it can be compactly written as
\begin{equation}
a^x \, {\rm mod} \, N \rightarrow x \, {\rm mod} \, r(a).
\end{equation}
Then, after changing the initial state of the second register from $|00 \cdots 1\rangle$ to $|00 \cdots 0 \rangle$, we have, instead of (\ref{modular exponentiation operator}), the compressed modular exponentiation operation 
\begin{equation}
| x \rangle \otimes |0 \cdots 0 \rangle \rightarrow | x \rangle \otimes | x \, {\rm mod} \, r \rangle.
\label{compressed modular exponentiation operator}
\end{equation}
The operation (\ref{compressed modular exponentiation operator}) without the modulo $r$ is just the bit-wise {\sf COPY} shown in Fig.~\ref{copy circuit figure}, and the effect of the modulo $r$ is to only copy the $\log_2 r$ least significant bits.

In conclusion, we require $\ell_{\rm max}$ qubits in each register, for a total of $2 \ell_{\rm max}$ qubits. $\ell_{\rm max}$ can either be computed classically or the bound (\ref{max l condition}) can be used. We note that the space requirements can be further reduced by using iterative phase estimation \cite{MoscaLNCS99,ParkerPRL00,DobsicekPRA07}, but with an increase in circuit depth. This might be useful for ion-trap and optical realizations but probably not for superconducting qubits.

\section{Factoring 51 and 85}

In this section we provide explicit quantum circuits for the cases of $N=51$ and $85$. In both cases $\ell_{\rm max} = 4$ (the largest order is 16), so we require $n=4$ qubits in the first register and $m=4$ in the second, for a total of 8 qubits. This is significantly fewer than the $3b$ required for general $b$-bit numbers ($b=6$ when $N\!=\!51$ and $b=7$ when $N\!=\!85$). It is also fewer than the $2b+3$ qubits required by Beauregard \cite{BeauregardQIC03}.

\begin{widetext}

\begin{figure}
\includegraphics[width=13.0cm]{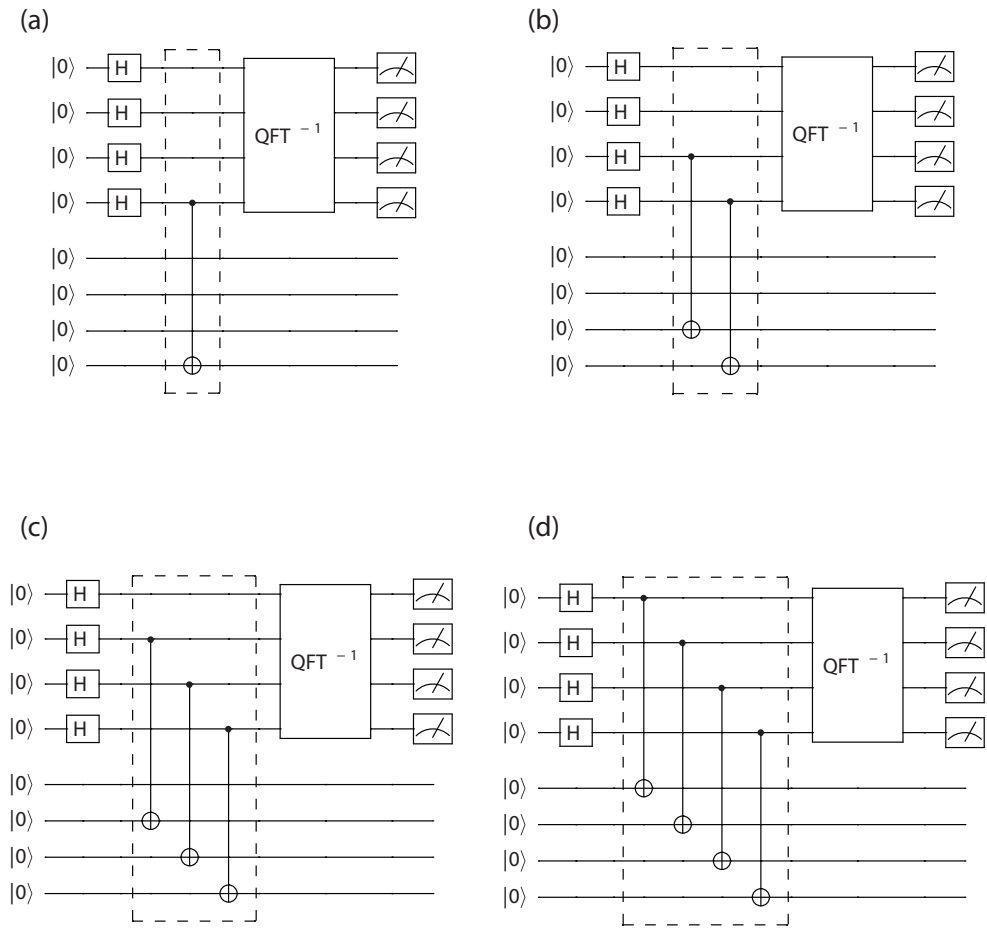} 
\caption{Quantum circuits for factoring $51$ and $85$. Note the modification of the 
input to the last qubit of the second register compared with Fig.~\ref{basic circuit figure}.
The circuits inside dashed boxes are the compressed modular exponentiation operations
discussed in Sec.~\ref{space requirements and circuit construction section}. Note that the
{\sf CNOT} gates here can be executed in parallel.}
\label{quantum circuit figure}
\end{figure} 

\end{widetext}

After the compression discussed in Sec.~\ref{space requirements and circuit construction section}, only four different circuits are needed to cover all $N\!=\!51$ and $N\!=\!85$ cases, because there are four possible orders. The assignments are listed in Tables \ref{51 table} and \ref{85 table}, and the circuits are given in Figs.~\ref{quantum circuit figure}a-d.

\begin{table}[htb]
\centering
\caption{\label{51 table} $N\!=\!51$ quantum circuits. The base marked by an asterisk 
satisfies $a^{r/2} \! = \! -1 \, {\rm mod} \, N$ and will result in a factorization failure in the 
classical post-processing analysis.}
\begin{tabular}{|c|c|}
\hline
base $a$ & circuit \\
\hline 
16, 35, 50$^*$   & Fig.~\ref{quantum circuit figure}a \\
\hline 
4, 13, 38, 47  & Fig.~\ref{quantum circuit figure}b \\
\hline 
2, 8, 19, 25, 26, 32, 43, 49   & Fig.~\ref{quantum circuit figure}c \\
\hline 
5, 7, 10, 11, 14, 20, 22, 23, 28, 29, 31, 37, 40, 41, 44, 46   & Fig.~\ref{quantum circuit figure}d \\
\hline 
\end{tabular}
\end{table}

\begin{table}[htb]
\centering
\caption{\label{85 table} $N\!=\!85$ quantum circuits. Bases marked by an asterisk satisfy 
$a^{r/2} \! = \! -1 \, {\rm mod} \, N$ and result in factorization failures in the classical 
post-processing analysis.}
\begin{tabular}{|c|c|}
\hline
base $a$ &  circuit \\
\hline 
16, 69, 84$^*$  & Fig.~\ref{quantum circuit figure}a \\
\hline 
4, 13$^*$, 18, 21, 33, 38$^*$, 47$^*$, 52, 64, 67, 72$^*$, 81  & Fig.~\ref{quantum circuit figure}b \\
\hline 
2, 8, 9, 19, 26, 32, 36, 42, 43, 49, 53, 59, 66, 76, 77, 83 & Fig.~\ref{quantum circuit figure}c \\
\hline 
 3, 6, 7, 11, 12, 14, 22, 23, 24, 27, 28, 29, 31, 37, 39, 41,   & \\
44, 46, 48, 54, 56, 57, 58, 61, 62, 63, 71, 73, 74, 78, 79, 82  & Fig.~\ref{quantum circuit figure}d \\
\hline 
\end{tabular}
\end{table}

\section{Conclusions}
\label{conclusion section}

Given the considerable interest in experimental demonstrations of Shor's algorithm, it is reasonable to ask what constitutes a ``genuine" demonstration of this important algorithm, and whether the cases presented here should be considered as such. In our opinion a genuine implementation should use no knowledge of the value of the order $r$---including whether or not it is a power of two---because the objective of the quantum stage of the algorithm is to calculate $r$. Therefore we do not regard the factorization of products of Fermat primes to be genuine implementations of Shor's algorithm. Moreover, such special cases can be efficiently factored classically, by comparing $N$ against a list of products of these primes.

However we do view the circuits presented here as quasi-legitimate implementations of quantum order finding, and in our view they are still interesting for this reason \cite{usingOrderNote}. In particular, each eight-qubit circuit presented here is able to detect periods of two, four, eight, and sixteen, so there is a failure mode where an incorrect period could be observed. But these genuine order-finding instances are nongeneric cases from the perspective of Shor's algorthm.

Smolin, Smith, and Vargo \cite{SmolinPre13} recently addressed the question of what should constitute a genuine factoring demonstration by simplifying the entire order-finding circuit for any product of distinct odd primes down to only {\it two} qubits. This is possible by implementing the phase estimation iteratively \cite{MoscaLNCS99,ParkerPRL00,DobsicekPRA07} (or the Fourier transform semiclassically \cite{GriffithsPRL96}), and by choosing only bases $a$ with order two. Smolin {\it et al.}~\cite{SmolinPre13} show that {\it with knowledge of the factors}, it is always possible to find an order-two base, and provide an algorithm for doing so. The circuit of Smolin {\it et al.}~does not constitute a genuine implementation of Shor's algorithm either. However the focus of our work is different than Ref.~\cite{SmolinPre13}, as the circuits presented here are still quasi-legitimate implementations of order finding, and we do not make explicit use of the factors in simplifying the circuits.

\begin{figure}
\includegraphics[width=7.0cm]{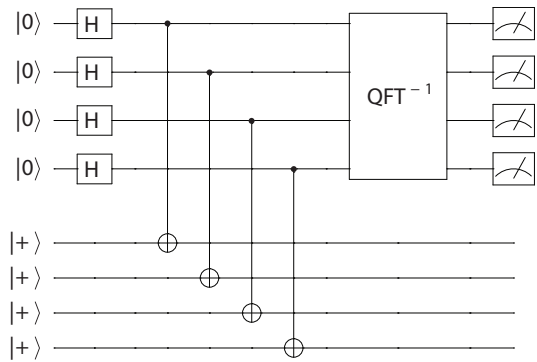} 
\caption{Changing the input states on the second register to verify coherent operation of the {\sf CNOT} gates.}
\label{plus initial state figure}
\end{figure}

Finally, we note that the $r=16$ cases (Fig.~\ref{quantum circuit figure}d) result in a uniform probability distribution for observing computational basis states $|x\rangle$ after measurement of the first register, which would also result from an unintended, purely decohering action of the {\sf CNOT} gates \cite{uniformPeaksNote}. One method of verifying that the circuit is functioning correctly is to perform tomography on the final state. A simpler method, however, is to change the input of the second register from $|0\rangle^{\otimes 4}$ to $|+\rangle^{\otimes 4}$, as shown in Fig.~\ref{plus initial state figure}. If the gates are purely decohering, this will not change the output of the first register upon measurement. But if the {\sf CNOT}s are acting ideally, the entire compressed modular exponentation operator now acts as the identity [because $|+\rangle\equiv 2^{-1/2}(|0\rangle+|1\rangle)$ is an eigenvector of the {\sf NOT} gate] and can be effectively dropped from the circuit, leading to an observation of the final state $|0000\rangle$ with unit probability.

In conclusion, we have shown that the simple and well-studied case of factoring $N\! =\!15$ is
the first in a series of cases 
\begin{equation}
15, 51, 85, 771, 1285, 4369, \dots
\end{equation}
that have all orders equal to a power of two and that can be factored with fewer resources than that of other products with the same number of bits. 

\acknowledgements

This research was funded by the US Office of the Director of National Intelligence (ODNI), Intelligence Advanced Research Projects Activity (IARPA), through the US Army Research Office grant No.~W911NF-10-1-0334. All statements of fact, opinion, or conclusions contained herein are those of the authors and should not be construed as representing the official views or policies of IARPA, the ODNI, or the US Government. 

We thank Joydip Ghosh for useful discussions and for finding a mistake in
a previous version of this manuscript.

\bibliography{/Users/mgeller/Desktop/group/publications/bibliographies/MRGpre,/Users/mgeller/Desktop/group/publications/bibliographies/MRGbooks,/Users/mgeller/Desktop/group/publications/bibliographies/MRGgroup,/Users/mgeller/Desktop/group/publications/bibliographies/MRGqc-josephson,/Users/mgeller/Desktop/group/publications/bibliographies/MRGqc-architectures,/Users/mgeller/Desktop/group/publications/bibliographies/MRGqc-general,/Users/mgeller/Desktop/group/publications/bibliographies/MRGqc-TSC,/Users/mgeller/Desktop/group/publications/bibliographies/MRGqc-algorithms,endnotes}

\begin{thebibliography}{19}
\bibitem [{usi()}]{usingOrderNote}%
  \BibitemOpen
  \href@noop {} {}\bibinfo {note} {In this work we simplify the modular
  exponentiation circuits to reduce their depth. However it is also possible to
  implement uncompiled versions, which do not make any use of the value of $r$
  and which would constitute a fully genuine implementation of order-finding
  (but not of factoring). The main point of this work, that the number of
  qubits required in the first register is greatly reduced for composites in
  the series (\ref{fermat products}), applies to either approach.}\BibitemShut
  {Stop}%
\end{thebibliography}

\end{document}